\documentclass[prb,reprint,twocolumn,superscriptaddress,oneside,floatfix]{revtex4-1}

\usepackage{epsfig,epstopdf,multirow,amsmath,amssymb,mathrsfs}

\begin{document} 
\title{Theory for charge and orbital density-wave states in manganite La$_{0.5}$Sr$_{1.5}$MnO$_4$}

\author{Zi-Jian Yao} \affiliation{Department of Physics and Center of Theoretical and Computational Physics, The University of Hong Kong, Hong Kong, China}

\author{Wei-Qiang Chen} \affiliation{Department of Physics, South University of Science and Technology of China, Shenzhen 518055, China} \affiliation{Department of Physics and Center of Theoretical and Computational Physics, The University of Hong Kong, Hong Kong, China}

\author{Jin-Hua Gao} \affiliation{Department of Physics, Huazhong University of Science and Technology, Wuhan 430074, China}

\author{Hong-Min Jiang} \affiliation{Department of Physics, Hangzhou Normal University, Hangzhou 310036, China} \affiliation{Department of Physics and Center of Theoretical and Computational Physics, The University of Hong Kong, Hong Kong, China}

\author{Fu-Chun Zhang} \affiliation{Department of Physics and Center of Theoretical and Computational Physics, The University of Hong Kong, Hong Kong, China} 
\begin{abstract}
	We investigate the high temperature phase of layered manganites, and demonstrate that the charge-orbital phase transition without magnetic order in La$_{0.5}$Sr$_{1.5}$MnO$_4$ can be understood in terms of the density wave instability. The orbital ordering is found to be induced by the nesting between segments of Fermi surface with different orbital characters. The simultaneous charge and orbital orderings are elaborated with a mean field theory. The ordered orbitals are shown to be $d_{x^2-y^2} \pm d_{3z^2-r^2}$. 
\end{abstract}

\maketitle

\section{Introduction} The manganese oxides are prototype materials for the rich physics of the interplay among spin,
charge, and orbital degrees of freedom, which has been an important issue in correlated electron
systems~\cite{tokura:462,dagotto:1,dagotto:67}. Though intensive theoretical studies on the phase transitions of
manganites have been carried out, most of them concentrate on the ground state, where the kinetic energy is subject to the
static spin order hence the orbital and charge ordering may
emerge~\cite{brink:1016,brink:5118,solovyev:2825,efremov:853}. Nevertheless, in the single-layered perovskite
La$_{0.5}$Sr$_{1.5}$MnO$_4$, which we will focus on this paper, the spin and charge-orbital phase transitions are
separated. With decreasing temperature, before the antiferromagnetic spin ordering that emerges at $T = T_N \approx$ 110
K ~\cite{sternlieb:2169}, a charge-orbital ordering phase transition emerges at $T =T_{co} \approx$ 220
K~\cite{moritomo:3297,sternlieb:2169,murakami:1932}. The charge density has a checkerboard distribution, and the orbital
has an ordered wave vector $(\pi/2,\pi/2)$. Another observation which may put doubt on the relevance between magnetic
order and charge/orbital orderings is that although similar charge/orbital orderings are experimentally observed in
single layer and bilayer manganites, the intralayer magnetic ordering is antiferromagnetic for the former but ferromagnetic for the latter. To understand such phenomenon, it would be important to investigate the mechanism of charge and orbital ordering in the absence of spin order.

The physics of manganites is usually described by the strong coupling approaches. For undoped manganite LaSrMnO$_4$, the high temperature orbital ordering could be achieved from the strong coupling approach~\cite{daghofer:277}. And for the half-doped La$_{0.5}$Sr$_{1.5}$MnO$_4$, the low temperature phase transition has been studied previously~\cite{daghofer:104451}. But various angle-resolved photoemission spectroscopy (ARPES) experiments on different layered manganites suggest an essential connection between the Fermi surface (FS) nesting and the charge/orbital ordering in this family of materials. La$_{1-x}$Sr$_{1+x}$MnO$_4$ is insulating for all Sr concentrations $x$. However, the remnant FS of La$_{0.5}$Sr$_{1.5}$MnO$_4$, which is about 190 meV below the chemical potential, has been probed by ARPES~\cite{evtushinsky:147201}. The observed fermiology consists of a large hole-like FS around $(\pi,\pi)$ and a very small electron pocket around $(0,0)$. The segment of the hole-like FS is quite flat, which may induce good FS nesting and lead to charge and orbital orderings~\cite{evtushinsky:147201}. There are other ARPES experiments that also indicate nesting-induced charge/orbital ordering. In an early ARPES measurement of the bilayer manganite La$_{1.2}$Sr$_{1.8}$Mn$_2$O$_7$~\cite{chuang:1509}, the nesting wave-vector $(0.6\pi,0)$ is found to be consistent with the modulation vector observed by x-ray and neutron experiments~\cite{doloc:4393}. Another very recent ARPES measurement on bilayer manganite (La$_{1-z}$Pr$_z$)$_{1.2}$Sr$_{1.8}$Mn$_2$O$_7$ shows addition evidence of the FS nesting induced ordering, where the observed FSs are almost straight lines, and the nesting wave vectors $(\pi/2,0)$ is confirmed as a modulation vector above the ferromagnetic transition temperature by elastic high energy x-ray diffraction measurement~\cite{trinckauf:16403}. It will be beneficial to understand the underlying physics of the observed relation between the features of FS and charge/orbital orderings by investigating the high temperature charge-orbital phase transition from the weak-coupling approach.

In this paper, we focus on the high temperature charge-orbital phase transition of single-layer La$_{0.5}$Sr$_{1.5}$MnO$_4$. We propose that the basic physics of the high temperature phase and its phase transition may be understood in the large Hund's coupling limit, where the electronic structure is described by two-fold Mn-3d $e_g$-orbital electrons, whose spins are confined to be parallel to the local $t_{2g}$ spins~\cite{dagotto:1}. The transition to the charge and orbital ordered states is driven by FS nesting and the interactions between $e_g$ electrons, and can be examined by using mean field approximations. Our theory explains the simultaneous orbital and charge orderings in the single-layered La$_{0.5}$Sr$_{1.5}$MnO$_4$. The theory may also be applied to understand the experiments of the bilayer compounds La$_{1.2}$Sr$_{1.8}$Mn$_2$O$_7$~\cite{chuang:1509,mannella:438} and (La$_{1-z}$Pr$_z$)$_{1.2}$Sr$_{1.8}$Mn$_2$O$_7$~\cite{trinckauf:16403}.

\section{Model Hamiltonian}

We first consider the full interaction Hamiltonian, which is given by 
\begin{align}
	\label{fullham} \mathcal{H}_I =& U \sum_{i\alpha} n_{i\alpha\uparrow}n_{i\alpha\downarrow} + (U^{\prime}-\frac{1}{2}J) \sum_i n_{i1} n_{i2} \nonumber \\
	& - 2J \sum_{i} \vec {s}_{i1} \cdot \vec {s}_{i2} + J \sum_i c_{i1\uparrow}^{\dagger} c_{i1\downarrow}^{\dagger} c_{i2\downarrow} c_{i2\uparrow} \nonumber \\
	& - J \sum_{i\alpha} \vec {s}_{i\alpha} \cdot \vec {S}_i + V\sum_{<ij>}(n_{i1}+n_{i2})(n_{j1}+n_{j2}), \nonumber 
\end{align}
where $U$, $U^{\prime}$ are on-site intra- and inter- orbital direct Coulomb repulsive interactions, respectively, and $J>0$ the exchange Coulomb interaction or the Hund's rule coupling. By symmetry, $U=U'+2J$. $V$ is the nearest-neighbor (NN) site Coulomb interaction. $\vec {s}_{i\alpha}$ is the spin of an electron of orbital $\alpha$ at site $i$. We denote $\alpha =1$ for $d_{x^2-y^2}$ orbital and $\alpha=2$ for $d_{3r^2-z^2}$ orbital. $\vec {S}_i$ is the spin-$\frac{3}{2}$ of three localized $t_{2g}$ electrons at site i, and $n_{i\alpha}=n_{i\alpha\uparrow}+n_{i\alpha\downarrow}$ is the total electron number operator for a given orbital. In our model, the inter-orbital Coulomb repulsion between $e_g$ and $t_{2g}$ electrons is a constant, which can be absorbed into the chemical potential.

In the large Hund's coupling limit, where $U$, $U'$, $J$ are much larger than the kinetic energy term $H_0$ below, we shall assume, however, $U'-J$ to be comparable with the kinetic energy, and may even be treated as a perturbation from a technical point of view. We may argue for this limit that in a metallic phase, the Coulomb interaction $U'$ has a good screening, while the Hund's coupling J is not screened, so that $U'-J$ could be small. In this limit, we follow Ref.~\onlinecite{dagotto:1} to assume that $\vec {s}_{i\alpha}$ is parallel to $ \vec {S}_i$, and doubly occupied $e_g$ electrons on the same site is allowed because it costs an energy of $U'-J$. Since the local spin degrees of freedom of $e_g$ electrons are frozen, the $e_g$ electrons behave like spinless fermions. Note that the spin degrees of freedom of the $e_g$ electron is frozen only locally, and the spins at different Mn sites, hence the spins of $e_g$ electrons at different sites, may have different polarizations. $H_I$ in the large Hund's coupling limit then takes the form, 
\begin{equation}
	\label{H_I} H_I = U_0\sum_i n_{i1} n_{i2} + V\sum_{<ij>}(n_{i1}+n_{i2})(n_{j1}+n_{j2}), 
\end{equation}
with $U_0 = U^{\prime}-J$, and the spin polarization of the $e_g$ electrons is implied. 

The kinetic energy term of the $e_g$ electrons can be described by a NN hopping matrix of the two $e_g$-orbitals. The single particle part of the Hamiltonian reads, 
\begin{eqnarray}
	H_0 = -\sum_{\langle ij \rangle, \alpha, \beta} t^{\alpha, \beta}_{i\sigma_i, j\sigma_j} (c^{\dag}_{i\alpha \sigma_i}c_{j\beta \sigma_j} + H.c.), 
\end{eqnarray}
where $\sigma_i$ is the spin orientation of the $t_{2g}$ electrons at site $i$. The hopping integrals between the two sites depend on the relative spin orientation of the two spins~\cite{dagotto:1}. In the semi-classical limit, one will have $t^{\alpha, \beta}_{i\sigma_i, j\sigma_j} =\cos{(\frac{\theta_{ij}}{2})} t^{\alpha,\beta}_{ij}$ with $\theta_{ij}$ the relative angle of the two spins at sites $i$ and $j$~\cite{berryphase}.

The solution of $H_0$ strongly depends on the spin configurations of the localized $t_{2g}$ electrons. Here we consider a high temperature phase where the spins are random, and approximate $\cos{(\frac{\theta_{ij}}{2})} \approx \langle \cos{(\frac{\theta}{2})} \rangle$, which is an averaged value of the solid angle and is independent of the pair $\langle ij \rangle$. Then we have $t^{\alpha\beta}_{i\sigma_i, j\sigma_j} =\langle \cos{(\frac{\theta}{2})} \rangle t^{\alpha\beta}_{ij}$, and $H_0$ is reduced to a usual tight-binding model for spinless fermions~\cite{millis:5144,koller:104432}. The pre-factor $\langle \cos{(\frac{\theta}{2})} \rangle$ represents a reduction of the hopping integral due to the random spins~\cite{quasiparticle}. Note that the average value of $\cos{(\frac{\theta}{2})}$ in the solid angle space is 2/3. $H_0$ then can be written as, 
\begin{align}
	\label{h0} H_0 =& - \langle \cos{(\frac{\theta}{2})} \rangle \sum_{\vec k \alpha\beta} 2t^{\alpha\beta} \gamma_{\alpha\beta}({\vec k}) c^{\dagger}_{\vec k\alpha}c_{\vec k\beta}, 
\end{align}
where $t^{\alpha\beta}$ is the hopping integral along the $x$-axis. $\gamma_{11}=\gamma_{22}=\gamma_{+}$, $\gamma_{12}=\gamma_{21} =\gamma_{-}$, and $\gamma_{\pm}(\vec k) = \cos{k_x} \pm \cos{k_y}$. In what follows, we shall study $H = H_0 + H_I$ by solving $H_0$ first and studying the effect of $H_I$ in Eq.~\eqref{H_I} from a weak-coupling approach.

\begin{figure}
	\centering 
	\includegraphics[width=0.5 
	\textwidth]{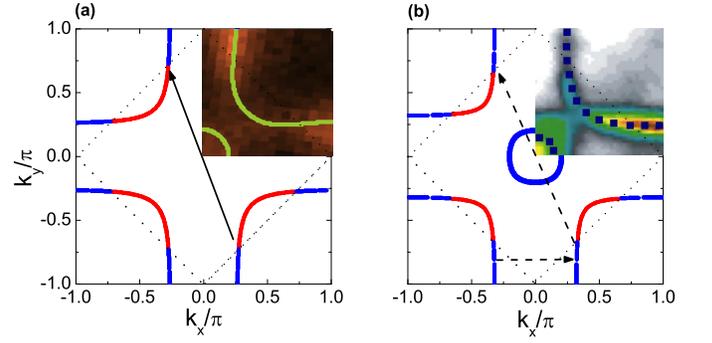} \caption{FS of the spinless fermion model $H_0$. (a) is for $e_g$ electron density n=0.5 per site, and (b) is for n=0.6. Red and blue colors on the Fermi sheets represent the states mostly orbital $d_{x^2-y^2}$ or $d_{3z^2-r^2}$, respectively. The upper-right $\frac{1}{4}$ BZ in (a) and (b) show the FS observed in the ARPES experiments on single-layered La$_{0.5}$Sr$_{1.5}$MnO$_4$~\cite{evtushinsky:147201} and on bilayer La$_{1.2}$Sr$_{1.8}$Mn$_2$O$_7$~\cite{mannella:438}, respectively. The colors used in ARPES data represent intensity, while the blue dots in (b) are added here to guide the eyes.} 
\end{figure}

$H_0$ can be diagonalized and the eigen-energy is given by 
\begin{align}
	\begin{split}
		\epsilon_{\pm} =& - \langle \cos{\theta/2} \rangle (t^{11} +t^{22})\gamma_{+}(\vec k) \nonumber \\
		& \pm \sqrt{(t^{11}-t^{22})^2\gamma_{+}(\vec k)^2 +4(t^{12})^2\gamma_{-}^2}. 
	\end{split}
\end{align}
The hopping matrix elements are related by Slater-Koster formalism~\cite{slater:1498} if we consider the direct hopping between the two NN Mn sites, from which we obtain $t^{22} = t^{11}/3$ and $t^{12} = t^{21} = -t^{11}/\sqrt{3}$. Hereafter, we will take $\langle \cos{\theta/2} \rangle t^{11}$ as the energy unit. 

Fig. 1(a) shows the calculated FS for the quarter filled $e_g$ electrons, namely 0.5 electron per Mn-site, relevant to the single layer La$_{0.5}$Sr$_{1.5}$MnO$_4$. As we can see, a large segment of the FS is quite flat, and there is a clear nesting at the wave vector $\vec q = (\pi/2, \pi/2)$, which suggests possible instabilities toward ordered states. Fig. 1(b) shows the FS for electron number 0.6 per Mn site, corresponding to the electron density of the bilayer compound La$_{1.2}$Sr$_{1.8}$Mn$_2$O$_7$~\cite{mannella:438}, where the bilayer splitting can be neglected. It is seen that the shape of the FS in each plot is in good agreement with the ARPES results.

\section{Orbital density-wave instability} 

We now study the effect of $H_I$. We will first identify the most plausible instabilities by using the random phase approximation (RPA) analysis. We then apply a mean field approach to examine the phase transitions. To study the density-wave instabilities, we define the following orbital (o) and charge (c) density operators, 
\begin{align}
	\rho^o_i &= n_{i+}-n_{i-}=c_{i1}^{\dagger}c_{i2}+c_{i2}^{\dagger}c_{i1}, \nonumber \\
	\rho^c_i &= n_{i+}+n_{i-}=c_{i1}^{\dagger}c_{i1}+c_{i2}^{\dagger}c_{i2}, 
\end{align}
where the orbitals $+$ and $-$ are linear combinations of the orbitals $d_{x^2-y^2}$ and $d_{3z^2-r^2}$, $c^{\dagger}_{i\pm}=\frac{1}{\sqrt{2}}(c^{\dagger}_{i1}{\pm}c^{\dagger}_{i2})$. As it will become clear later, the orbital ordering in this problem is associated with orbitals $+$ and $-$, instead of $1$ and $2$. We introduce a static susceptibility matrix $\hat{\chi}$, whose element is defined as 
\begin{equation}
	\chi_{\alpha \alpha^{\prime}, \mu^{\prime} \mu}(q) = \frac{1}{2} \int_0^{\beta} d\tau \left<T_{\tau} \rho_{\alpha\alpha'}(\vec q, \tau) \rho_{\mu\mu'}(-\vec q, 0) \right>, 
\end{equation}
where $\rho_{\alpha\alpha'}(\vec q) = \sum_{\vec k}c^{\dag}_{\vec k +\vec q, \alpha}c_{\vec k, \alpha'}$.

The orbital and charge susceptibilities are then given by 
\begin{eqnarray}
	\label{rf} \chi^o(\vec q) &= \frac{1}{2} \sum_{\alpha \mu} \chi_{\alpha \bar{\alpha}, \mu \bar{\mu}}(\vec q), \nonumber \\
	\chi^c(\vec q) &= \frac{1}{2} \sum_{\alpha \mu}\chi_{\alpha \alpha, \mu \mu}(\vec q). 
\end{eqnarray}
Within the RPA, we have $\hat{\chi} = (\hat{I}+\hat{\chi}^{(0)} \hat{U}^c )^{-1}\hat{\chi}^{(0)}$, where $\hat I$ is an identity operator, and $\hat{\chi}^{(0)}$ is the matrix of the bare susceptibility, 
\begin{eqnarray}
	\label{chi0} \chi^{(0)}_{\alpha \beta, \mu \nu}(\vec q) =\frac{1}{N}\sum_{\vec k mn}a^{\alpha*}_{m}(\vec k+ \vec q)a^{\beta}_{n}(\vec k) a^{\nu*}_{n}(\vec k)a^\mu_{m}(\vec k + \vec q)\nonumber \\
	\times [ f(\epsilon_{n}(\vec k+ \vec q))-f(\epsilon_{m}(\vec k)) ]/[\epsilon_{m}(\vec k)-\epsilon_{n}(\vec k+\vec q)+i\eta ], \nonumber
\end{eqnarray}
where $m$ and $n$ are the band indices, and $a^{\alpha}_m(\vec k) = \left \langle \alpha,\vec k | m,\vec k \right \rangle $ is the orbital weight. We arrange the matrix index from 1 to 4 as $(\alpha \beta)$ = (11), (22), (12), and (21). The interaction matrix $\hat{U}^c$ is of the form $\hat{U}^c = \hat{U}^1 \oplus \hat{U}^2$, where $\hat{U}^1=V(\vec q) \sigma_0 + (V(\vec q)+U_0) \sigma_1$, and $\hat{U}^2= -U_0 \sigma_0$ with $\sigma_0$ an identity matrix and $\sigma_1$ the first Pauli matrix.

In the matrix representation described above, the upper-left 2 by 2 block in $\hat{\chi}$ describes charge part and the lower-right block describes the orbital part, as we can see from Eqs.~\eqref{rf}. While $\hat U^c$ is block diagonal, $\chi^{(0)}(\vec q)$ is generally not block diagonal, so that the charge and orbital are coupled in the response functions. A special case is at $q_x=\pm q_y$, where the off-diagonal components of $\hat{\chi}$ vanishes due to the symmetry in the band structure~\cite{diagonal}, which makes the study of the instability at $\vec q =(\pi/2, \pi/2)$ and $\vec q =(\pi, \pi)$ simpler. In this case, the inter-orbital nesting connecting the FS segments with different orbital character, favors the $\pm$ orbital ordering. To illustrate this point, we define orbitals $c_{\alpha}^\dagger=\cos(\psi)c_1^\dagger+\sin(\psi)c_2^\dagger$ and $c_{\beta}^\dagger=\sin(\psi)c_1^\dagger-\cos(\psi)c_2^\dagger$, with orbital density $\rho^o_\psi=c_\alpha^{\dagger}c_\alpha-c_\beta^{\dagger}c_\beta$. In the vicinity of orbital-density-wave instability, because of the dominant role of inter-orbital nesting, we have $\chi^o_\psi \approx \cos(\psi)\sin(\psi) \sum_{\alpha \mu} \chi_{\alpha \bar{\alpha}, \mu \bar{\mu}}$, which reaches its maximum with $\psi=\pi/4$. Therefore the ordered orbitals are $+$ and $-$ (for a more detailed discussion, see Ref.~\onlinecite{yao:37009}). 
\begin{figure}
	\label{fig:2} \center \epsfig{figure=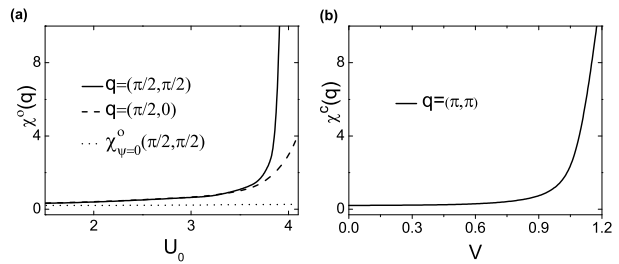,width=0.45 
	\textwidth} \caption{$T=0$ RPA susceptibilities. (a) Orbital susceptibility $\chi^o(\vec{q})$ at $\vec{q}=(\pi/2,\pi/2)$ and $(\pi/2,0)$ for orbital ordering between $+$ and $-$ with $V=0$. The dotted line is the susceptibility for orbital ordering between orbitals 1 and 2. (b) Charge susceptibility $\chi^c(\vec{q})$ at $\vec{q}=(\pi,\pi)$ with $U_0 = 0$.} 
\end{figure}

We have found three types of instabilities in our calculations, namely the orbital ordering at $(\pi/2,\pi/2)$ and at $(\pi/2,0)$, and the charge ordering at $(\pi,\pi)$. Note that the orbital orderings are related to the FS nesting, while the charge ordering is not. In Fig. 2 we plot the susceptibilities at corresponding wave vectors as functions of interaction strengths. As we can see, the orbital susceptibilities at $(\pi/2,\pi/2)$ and $(\pi/2,0)$ are greatly enhanced by the inter-orbital repulsion $U_0$, and the susceptibility at $(\pi/2,\pi/2)$ is much larger at large $U_0$ with a critical value of $U_0 \approx 4$ for the ordering. Note that the orbital susceptibility based on the ordering between orbitals $d_{x^2-y^2}$ and $d_{3z^2-r^2}$, $\chi_{\psi=0}^o(\pi/2,\pi/2)$, is much weaker as we can see from the dotted line in Fig. 2(a). For the charge ordering at $(\pi,\pi)$, as plotted in Fig. 2(b), $\chi^c$ diverges at $V \approx 1.1$, which indicates a phase transition to $(\pi,\pi)$ charge ordering.

The picture of nesting-induced density wave could also be applied to understand the ordering of the bilayer manganite La$_{1.2}$Sr$_{1.8}$Mn$_{1.2}$O$_7$ which has a ferromagnetic-metal ground state. Since the bilayer splitting is not observed in ARPES experiments~\cite{chuang:1509,trinckauf:16403}, we simply ignore it. The FS with orbital character is shown in Fig. 1(b). As seen there are basically two nesting wave vectors, the intra-orbital one is $q_1=(0.6\pi,0)$, and the inter-orbital one is $q_2=(0.6\pi,0.6\pi)$. It is claimed that $q_1$ is the charge-ordering wave vector~\cite{doloc:4393}, which is consistent with our understanding that intra-orbital nesting favors CDW. The nesting at $q_2$ should induce an orbital order, but so far there is no experimental evidence for this ordering. Interestingly, peaks of static susceptibility at wave vectors around $q_2$ are reported in a first principle study~\cite{saniz:236402}. 

\section{Phase diagram and phase transition} The RPA calculations above have indicated two possible major instabilities, the $(\pi/2,\pi/2)$ orbital order (OO) and $(\pi,\pi)$ charge order (CO). Below we use a mean field approach to examine the interplay between the two orderings. We introduce two mean fields 
\begin{align}
	\left\langle \rho_{i}^o \right\rangle & = \left\langle c_{i1}^{\dag} c_{i2} + c_{i2}^{\dag} c_{i1} \right\rangle = \rho_o \cos(\mathbf{q}_1 \cdot \mathbf{r}_i + \phi), \nonumber\\
	\left\langle \rho_i^c \right\rangle & = \left\langle c_{i1}^{\dag} c_{i1} + c_{i2}^{\dag} c_{i2} \right\rangle = \rho_c \cos(\mathbf{q}_2 \cdot \mathbf{r}_i) + \bar{\rho}, 
\end{align}
with $\mathbf{q}_1=(\pi/2,\pi/2)$, $\mathbf{q}_2=(\pi,\pi)$, and $\bar{\rho} = 0.5$. $\rho_o$ and $\rho_c$ are the order parameters of charge and orbital, respectively, while $\phi$ is the phase shift in the real space of the orbital order. The mean field Hamiltonian then reads 
\begin{align}
	\label{hmf} H_{\mathrm{MF}} & = H_0 - \frac{U_0}{4} \sum_{\mathbf{k}}\left[\rho_o e^{i \phi} (c_{\mathbf{k}, 2}^{\dag} c_{\mathbf{k}+\mathbf{q}_1, 1} + c_{\mathbf{k}, 1}^{\dag} c_{\mathbf{k+q_1}, 2}) + H.c. \right] \nonumber\\
	& \phantom{=} - 4(V-U_0/8) \rho_c \sum_{\mathbf{k} \alpha} c_{\mathbf{k}, \alpha}^{\dag} c_{\mathbf{k}+\mathbf{q_2}, \alpha}. 
\end{align}
The self consistent equations for the mean fields are 
\begin{align}
	\label{eq:1} \rho_o e^{i \phi} & = \frac{2}{N} \sum_{\mathbf{k}} \left\langle c_{\mathbf{k+q_1}, 1}^{\dag} c_{\mathbf{k},2} + c_{\mathbf{k+q_1},2}^{\dag} c_{\mathbf{k},1} \right\rangle, \nonumber\\
	\rho_c & = \frac{1}{N} \sum_{\mathbf{k} \alpha} \left\langle c_{\mathbf{k+\mathbf{q}_2}, \alpha}^{\dag} c_{\mathbf{k},\alpha} \right\rangle. 
\end{align}
\begin{figure}
	\centerline{ 
	\includegraphics[width=0.5 
	\textwidth]{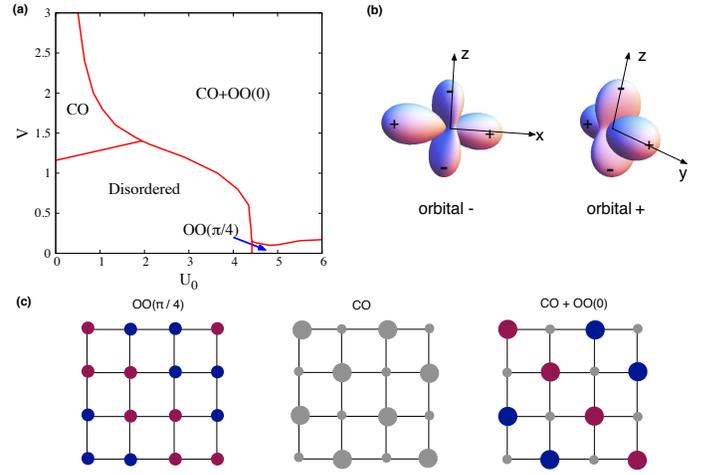}} \caption[]{\label{fig:phase_diagram} (a) Phase diagram at zero temperature. CO: charge-ordered phase; OO: orbital-ordered phase. (b) Shapes of ordered orbitals $+$ and $-$. (c) Illustration of the ordered states in real space. Electron charge is represented by the size of the circle. Orbitals are represented by colors: Blue for dominant orbital +, green for dominant orbital -, and grey for orbital-disordered site.} 
\end{figure}

By solving $H_{\mathrm{MF}}$ together with the self-consistent equations \eqref{eq:1}, we obtain the zero temperature phase diagram, which is shown in Fig.~\ref{fig:phase_diagram}. In the calculation, we found only two possible phase shift $\phi$ for the orbital ordering, $\phi=\pi/4$ and $\phi=0$, which are denoted as OO$(\pi/4)$ and OO(0), respectively, in the phase diagram. The real space modulation of each phase is sketched in Fig.~\ref{fig:phase_diagram}(c). One of the main features of the phase diagram is that the system is in the co-existence phase of CO and OO(0) in a large parameter space of $(U_0, V)$. The phase with just the orbital ordering appears in a tiny phase space with very small $V$ and large $U_0$. We also note that there is a sudden change on the orbital ordering phase from OO$(\pi/4)$ in the absence of CO to OO(0) in the presence of CO. Below we shall provide some understanding of the latter. Let us first consider the orbital ordered only phase. The preferred phase OO$(\pi/4)$ may be understood as the result of losing less kinetic energy due to the orbital ordering. The amplitude of the orbital order parameter $\left\langle \rho_{i}^o \right\rangle$ for the OO$(\pi/4)$ phase is $\rho_o / \sqrt{2}$, while the amplitude for the OO(0) phase is $\rho_o$. However, the situation is very different in the presence of charge ordering. In that case, the local orbital order $\left\langle \rho_{i}^o \right\rangle$ is bound by the local charge density of electrons $\left\langle \rho_i \right\rangle$. Because $\left\langle \rho_i \right\rangle$ are reduced on some sites, the charge ordering suppresses the OO$(\pi/4)$ phase. On the other hand, the OO(0) is consistent with and may even be enhanced by the charge ordering. In the limit of strong charge ordering $\rho_c = 1/4$, the kinetic energy term diminishes, and $\left\langle \rho_o \right\rangle =1$ in the phase OO(0), in comparison with a maximum value of $\left\langle \rho_o \right\rangle = 0.5$ in the absence of charge ordering. In other words, the presence of charge order will induce the OO$(0)$ phase. The transition from OO$(\pi/4)$ phase to CO+OO(0) phase is the first order.

In Fig. 4(a)-(c), we plot the orbital and charge order parameters as functions of $V$ for various $U_0$ at $T=0$. At small $U_0 =1$, as $V$ increases, CO develops first followed by a co-existent phase with the OO(0) order. At $U_0=3$, the transition to the charge and orbital ordered state is simultaneous as $V$ increases, and is first order with clear jumps in the order parameters. At large $U_0=5.5$, we have only orbital ordering at small $V$, and co-existent phase with charge ordering. And at the charge ordering point, the orbital order parameter has a change in both the phase (not shown here but discussed before) and its magnitude. In Fig. 4(d), we show the order parameters as functions of temperature for $(U_0=4, V=1)$, to illustrate the simultaneous first order phase transition of the orbital and charge orderings at finite temperature~\cite{band_renorm}, which may explain the simultaneous orderings observed in experiment of La$_{0.5}$Sr$_{1.5}$MnO$_4$~\cite{murakami:1932}. 
\begin{figure}
	\centerline{ 
	\includegraphics[width=0.45 
	\textwidth]{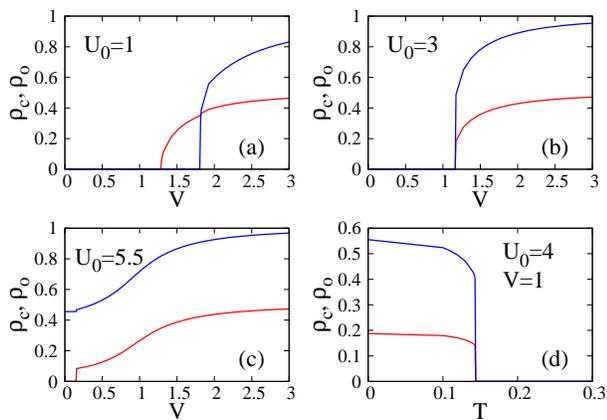}} \caption[]{\label{fig:transition} Panels (a), (b), and (c): $V$ dependence of orbital ordering $\rho_o$ (blue curves) and charge ordering $\rho_c$ (red curves) (a) at $U_0=0$; (b) at $U_0 = 3$; (c) at $U_0 = 5.5$. Panel (d): Temperature dependence of orbital ordering $\rho_o$ (blue curves) and charge ordering $\rho_c$ (red curves) at $U_0 = 4, V = 0.5$.} 
\end{figure}

We now discuss the ordered orbital characters of the single-layered system. Different from the usual rotational
invariant spin-1/2 space, the kinetic energy term is not symmetric with respect to the rotation in the pseudo-spin $e_g$
orbital space. Therefore, there is a selection of specific orbitals for the orbital-density-wave ordering. The ordered
orbitals have been suggested to be $d_{3x^2-z^2}$ and
$d_{3y^2-z^2}$~\cite{daghofer:104451,mirone:23,wu:155126}. Meanwhile, some x-ray scattering
experiments~\cite{wilkins:167205,huang:87202} combined with local-density approximation including on-site Coulomb
interactions (LDA+U) calculations~\cite{huang:87202} indicate that the orbital ordering is dominated by $d_{x^2-z^2}$
and $d_{y^2-z^2}$, which is also supported by means of x-ray structural analyses~\cite{okuyama:64402}. To further
examine this issue, we have performed the mean field calculations to examine the ordering between a general linear
combination of $d_{3z^2-r^2}$ and $d_{x^2-y^2}$, and have found that the ordering between $+$ and $-$ has the lowest
energy, which is also consistent with our RPA analysis. Therefore, in contrast to previous arguments that the ordered
orbitals are non-orthogonal, we propose that at temperature $T_N \le T \le T_{co}$, the ordered orbitals are orthogonal
orbitals $+$ and $-$, which are essentially equal mixtures of $d_{3z^2-r^2}$ and $d_{x^2-y^2}$. The shape of each
orbital is plotted in Fig. 3(b). We note that although our theory is qualitative, the ordered orbitals $+$ and $-$ are
actually selected by the symmetry of Hamiltonian.  Our results may provide a guideline for further study of more refined
numerical approaches such as Monte Carlo simulation~\cite{yunoki:3714,cen:97203} and density functional theory
calculations~\cite{huang:87202,trinckauf:16403}.

\section{summary} 

In summary, we have proposed that the basic physics of the high temperature phase in layered manganite La$_{0.5}$Sr$_{1.5}$MnO$_4$ may be described by an effective band Hamiltonian. Our theory reveals the essential connection between FS nesting and charge/orbital ordering, and explains the simultaneous phase transition to the charge and orbital ordered state. 

\begin{acknowledgments}
	Part of the work is supported by Hong Kong RGC grant HKU707010. 
\end{acknowledgments}

\bibliographystyle{apsrev} 
\bibliography{ref} 
\begin{thebibliography}{33}
\bibitem[{qua()}]{quasiparticle}
\bibinfo{note}{The broadening in the single particle energy spectrum introduced
  by randomized effective hopping integrals is much narrower than all of the
  relevant energy scales, such as the Fermi energy and the interaction
  strength. Therefore, it is a good approximation to renormalize the bare
  hopping integrals with the average value of the relative orientation of the
  local $t_{2g}$ spins.}
\bibitem[{dia()}]{diagonal}
\bibinfo{note}{The block off-diagonal index can generally be written to be
  $(\alpha, \alpha, \alpha, \bar{\alpha})$ and its cyclic. Under the exchange
  of $k_x$ and $k_y$, we have dispersion $\epsilon_{\pm}(k_x, k_y)
  =\epsilon_{\pm}(k_y, k_x)$, and the orbital weight $a^{\alpha}_{n}(k_x, k_y)
  a^{\beta*}_{n}(k_x, k_y) = (-1)^{\alpha+\beta} a^{\alpha}_{n}(k_y,k_x)
  a^{\beta*}_{n}(k_y, k_x)$. Therefore at $q_x =q_y$, we have
  $\chi^{(0)}_{\alpha \alpha, \alpha \bar{\alpha}}(q_x, q_y) =
  -\chi^{(0)}_{\alpha \alpha, \alpha \bar{\alpha}}(q_y, q_x)=0$. The same
  analysis can be applied to show that the other block off-diagonal elements
  vanish at $\vec q =(q_x, \pm q_x)$.}
\bibitem[{ban()}]{band_renorm}
\bibinfo{note}{Note that the mean field theory predicts a large first order
  transition, which may change the electronic structure dramatically under the
  transition. Since mean field theories usually over-estimate the order
  parameter jumps in the first order phase transition, the interpretation of
  this should be more cautious.}
\end{thebibliography}
\end{document}